\documentclass[10pt,preprint]{aastex}

\shorttitle{Cohesion of Amorphous Silica Spheres}
\shortauthors{Kimura et al.}

\begin{document}


\title{COHESION OF AMORPHOUS SILICA SPHERES: TOWARD A BETTER UNDERSTANDING OF THE COAGULATION GROWTH OF SILICATE DUST AGGREGATES}


\author{\sc Hiroshi Kimura\altaffilmark{1,2}}
\affil{Graduate School of Science, Kobe University, c/o CPS (Center for Planetary Science), \\
    Chuo-ku Minatojima Minamimachi 7-1-48, Kobe 650-0047, Japan}
\email{hiroshi\_kimura@cps-jp.org}

\author{\sc Koji Wada\altaffilmark{2}}

\author{\sc Hiroki Senshu\altaffilmark{2}}
\affil{Planetary Exploration Research Center (PERC), Chiba Institute of Technology (Chitech), \\Tsudanuma 2-17-1, Narashino, Chiba 275-0016, Japan}

\and

\author{\sc Hiroshi Kobayashi}
\affil{Department of Physics, Nagoya University, Chikusa-ku Furo-cho, Nagoya 464-8602, Japan}


\altaffiltext{1}{CPS Visiting Scientist}
\altaffiltext{2}{CPS Collaborating Scientist}


\begin{abstract}
Adhesion forces between submicrometer-sized silicate grains play a crucial role in the formation of silicate dust agglomerates, rocky planetesimals, and terrestrial planets.
The surface energy of silicate dust particles is the key to their adhesion and rolling forces in a theoretical model based on the contact mechanics.
Here we revisit the cohesion of amorphous silica spheres by compiling available data on the surface energy for hydrophilic amorphous silica in various circumstances. 
It turned out that the surface energy for hydrophilic amorphous silica in a vacuum is a factor of 10 higher than previously assumed.
Therefore, the previous theoretical models underestimated the critical velocity for the sticking of amorphous silica spheres, as well as the rolling friction forces between them.
With the most plausible value of the surface energy for amorphous silica spheres, theoretical models based on the contact mechanics are in harmony with laboratory experiments.
Consequently, we conclude that silicate grains with a radius of $0.1~\mu$m could grow to planetesimals via coagulation in a protoplanetary disk.
We argue that the coagulation growth of silicate grains in a molecular cloud is advanced either by organic mantles rather than icy mantles or, if there are no mantles, by nanometer-sized grain radius.
\end{abstract}





\section{INTRODUCTION}

It is of great importance in astrophysics to understand the outcomes of mutual collisions between submicrometer-sized silicate grains as well as agglomerates of them at relative velocities of $1$--$100~\mathrm{m\,s^{-1}}$.
In their pioneering work, \citet{poppe-et-al2000} presented collision experiments in which amorphous silica spheres (Monospher$^{\textregistered}$, Merck) of radius $a=0.6~\micron$ stuck to an amorphous silica target unless the collision velocity exceeded 1.1--$1.3~\mathrm{m~s}^{-1}$.
However, this threshold velocity is one order of magnitude higher than the critical velocity for the sticking of the spheres that was expected from a theoretical consideration \citep[see][]{chokshi-et-al1993,dominik-tielens1997}.
\citet{blum-wurm2000} experimentally demonstrated that the energy necessary for breaking an existing contact between amorphous silica spheres is underestimated in the theoretical arguments \citep[see][for more experimental works]{guettler-et-al2010}. 
The theoretical estimates rely on the contact mechanics, in particular the so-called JKR theory of adhesive contact of \citet{johnson-et-al1971}.
\citet{schraepler-blum2011} measured the erosion rates of agglomerates consisting of nonporous amorphous silica spheres (sicastar$^{\textregistered}$, micromod Partikeltechnologie GmbH) with a radius of $a=0.76~\micron$ using collisions of the same silica spheres.
Owing to the discrepancy between experiments and models for the sticking of amorphous silica spheres, the measured erosion rates are much smaller than expected by theoretical models.
As a result, a modification of the JKR theory involving introducing additional energy dissipation has been a common practice for compensating for the discrepancy \citep{brilliantov-et-al2007,paszun-dominik2008,seizinger-et-al2013}.
It is most unfortunate that the majority of recent models tend to do nothing more than to focus on how to tune the energy dissipation in the models to experimental results on collisions.

According to the JKR theory, the critical velocity for the sticking of spherical particles is proportional to $\gamma^{5/6}$, where $\gamma$ denotes the surface energy of the particles \citep{chokshi-et-al1993,dominik-tielens1997}.
Following an experimental work by \citet{kendall-et-al1987a} or \citet{heim-et-al1999}, the surface energy of hydrophilic amorphous silica is commonly assumed to be $\gamma = 0.020$--$0.025~\mathrm{J~m}^{-2}$ in the previous models \citep{chokshi-et-al1993,dominik-tielens1997,wada-et-al2007,ringl-et-al2012,seizinger-et-al2013}.
These values for amorphous silica are smaller than those for H$_2$O ice, while substances with high melting temperatures have long been well-known to have high surface energies \citep[e.g.,][]{reynolds-et-al1976,israelachvili2011,yamamoto-et-al2014}\footnote{To be precise, the surface energy is, in theory, proportional to the melting temperature divided by the square of the nearest-neighbor separation \citep{reynolds-et-al1976,blairs-joasoo1981}. 
Thus, the conclusion of \citet{yamamoto-et-al2014}, who claim the proportionality of the surface energy to the melting temperature is flawed.}.
It is worthwhile to note that the surface chemistry of the particles, and hence their circumstances, play an important role in the determination of the surface energy \citep[e.g.,][]{parks1984,walton2007,israelachvili2011}.
In fact, it has been suggested that the surface energy of silica glass measured in a vacuum is one order of magnitude higher than that under ambient conditions \citep{orowan1944}.
\citet{kendall-et-al1987a} and \citet{heim-et-al1999} presented their results on the surface energy under ambient conditions, while \citet{poppe-et-al2000}, \citet{blum-wurm2000}, and \citet{schraepler-blum2011} performed their collision experiments in a vacuum.
The surfaces of hydrophilic amorphous silica are occupied by silanol (Si--OH) groups with at most a single monolayer of adsorbed H$_2$O molecules ($\theta_\mathrm{H_2O} \le 1$) in a vacuum \citep{zhuravlev2000}.
In contrast, hydrophilic amorphous silica surfaces under ambient conditions are found to swell by 1--2~nm, which reflect multiple layers of adsorbed H$_2$O molecules ($\theta_\mathrm{H_2O} > 1$) \citep{vigil-et-al1994}.
Therefore, when evaluating the value of $\gamma$ that is appropriate for mutual collisions of amorphous silica grains, one has to properly take into account the surface chemistry of interfering grain surfaces.

On the one hand, the hydrogen-bonded linkage at the outermost layer of adsorbed water is the main contributor for the surface energy measured by \citet{kendall-et-al1987a} and \citet{heim-et-al1999} at the equilibrium elastic displacements expected in the JKR theory.
On the other hand, hydrogen bonding between silanol groups may contribute to the cohesive force on hydrophilic silicate grains upon collision in a vacuum.
Because hydrogen bonding between silanol groups is stronger than hydrogen bonding between water molecules, it is likely that the previous models underestimated the surface energy for amorphous silica in collision experiments.
Consequently, we may not need additional energy dissipation to describe collision experiments in the framework of the JKR theory if we pay great attention as to which value of $\gamma$ is more appropriate for collision experiments.

\section{SURFACE ENERGY OF AMORPHOUS SILICA}

The surface energy $\gamma$ is the most important parameter for modeling mutual collisions between silicate grains, though its determination is not trivial.
First, we shall review a possible range of surface energies determined for amorphous silica surfaces under different conditions and techniques.
By summarizing the available data on the surface energies, we propose the most probable value for the surface energy of amorphous silica appropriate for collision experiments.

\subsection{Contact Spot Measurements}

\citet{kendall-et-al1987a,kendall-et-al1987b} determined the equilibrium diameter of contact between fine powders of amorphous silica spheres from an elastic modulus of the powder assembly.
They estimated the surface energy of $\gamma = 0.025~\mathrm{J~m}^{-2}$ for amorphous silica (AEROSIL$^{\textregistered}$ OX50, Degussa) with a silanol density of $\alpha_\mathrm{OH} =  2.0~\mathrm{OH/nm^2}$ at room temperature.
Because the elastic modulus increased with temperature, but did not change with immersion in water, they concluded that the surface of powder particles was saturated with atmospheric contaminations mainly of adsorbed water.

\subsection{Direct Pull-off Force Measurements}

\citet{bradley1932} measured pull-off forces between quartz glass spheres, one of which is supported by a spiral spring, while the other is attached to a rod.
His experimental data on the relation between the forces and the effective diameter of the spheres in air yield $\gamma = 0.045~\mathrm{J~m}^{-2}$ based on the JKR theory \citep{kendall2001}.

\subsection{Colloidal Probe Techniques}

In colloidal probe techniques, amorphous silica particles are glued to the cantilever of an atomic force microscope (AFM)  using an adhesive such as epoxy resin because they cannot be practically attached to a cantilever by sintering.
Using an AFM, \citet{fuji-et-al1999} measured the surface energies of nonporous hydrophilic amorphous silica particles (Admafine) with a radius $a=0.85~\micron$ and hydrophobic amorphous silica particles.
The hydrophobic amorphous silica particles were prepared by modification with hexamethyldisilazane (HMDS) to the surfaces of the hydrophilic amorphous silica particles, in which the outermost hydrogen atoms were partly replaced by trimethylsilyl (TMS) groups.
The surface energy of hydrophilic silica particles with $\alpha_\mathrm{OH} =  3.6$--$4.6~\mathrm{OH/nm^2}$ lies in the range of $\gamma = 0.020$--$0.032~\mathrm{J~m}^{-2}$ at relative humidities of 40--70\% and increases up to $\gamma = 0.055~\mathrm{J~m}^{-2}$ at relative humidities close to 90\%.
\citet{fuji-et-al1999} found values as small as $\gamma = 0.002~\mathrm{J~m}^{-2}$ for the highest TMS density of $1.16~\mathrm{TMS/nm^2}$, while the value increases with relative humidities.
This extremely low surface energy results from the shielding of surface silanol groups by the TMS groups from forming hydrogen bonds \citep{fuji-et-al1999}.
The surface energies for silica particles with the TMS density less than $1.0~\mathrm{TMS/nm^2}$ were $\gamma \approx 0.023~\mathrm{J~m}^{-2}$ at relative humidities 40--80\%, similar to the values for the original silica particles.
They claimed that their experiments were limited to relative humidities at 40\% and above, because the effect of tribological charging on their measurements cannot be neglected at relative humidities lower than 40\%.
A similar AFM measurement by \citet{heim-et-al1999} gave $\gamma = 0.0186 \pm 0.0032~\mathrm{J~m}^{-2}$ for nonporous, hydrophilic silica microspheres (Bangs Laboratories, Inc.) of $a=0.5$--$2.5~\micron$, independent of relative humidity in the range of 10--40\%.
A pull-off force for the same silica spheres (Bangs Laboratories, Inc.) of $a=1.5$--$2.5~\micron$ on an oxidized silicon wafer was measured by \citet{ecke-et-al2001}.
They determined $\gamma = 0.0307~\mathrm{J~m}^{-2}$ for the original surfaces and $\gamma = 0.004~\mathrm{J~m}^{-2}$ for silanated surfaces modified with HMDS, with a standard deviation of 30--40\%.
The identical experimental setup was used to measure sliding frictional forces between silica spheres (Bangs Laboratories, Inc.) and the oxidized silicon wafers \citep{ecke-butt2001}.
The surface energies for the original and silanated surfaces were $\gamma = 0.036~\mathrm{J~m}^{-2}$ and  $\gamma = 0.0055~\mathrm{J~m}^{-2}$, respectively, based on the JKR theory.
\citet{ling-et-al2007} measured a pull-off force between the same spheres (Bangs Laboratories, Inc.) of $a=1.4$--$2.1~\micron$ as \citet{heim-et-al1999} at relative humidities of $< 30\%$, with special care given for contamination. 
The measured forces correspond to the surface energies of $\gamma = 0.0425 \pm 0.0053~\mathrm{J~m}^{-2}$ for the original spheres and $\gamma = 0.0453 \pm 0.0148~\mathrm{J~m}^{-2}$ for plasma-treated spheres.

\subsection{Surface Forces Apparatus (SFA) Measurements}

A pull-off force measured between silica sheets mounted on an SFA in dry nitrogen indicates $\gamma = 0.052 \pm 0.003~\mathrm{J~m}^{-2}$ for silica glass (Suprasil$^{\textregistered}$, Heraeus), whose surface has a low degree of hydroxylation but is still hydrophilic \citep{horn-et-al1989}.
\citet{vigil-et-al1994} measured a pull-off force between amorphous silica films using an SFA and suggests $\gamma = 0.056~\mathrm{J~m}^{-2} (t/10~\mathrm{s})^{1/n}$ in humid air with 33\% relative humidity and $\gamma = 0.006~\mathrm{J~m}^{-2} (t/10~\mathrm{s})^{1/n}$ in dry air where $t$ indicates the time after contact and $n=8 \pm 1$.
The low surface energy of silica films in dry air results from the small size of the capillary neck formed by a low amount of water molecules.
They also derived $\gamma = 0.005$--$0.015~\mathrm{J~m}^{-2}$ from the equilibrium diameter of the contact spot using friction forces between amorphous silica films measured in dry air, while $\gamma = 0.071 \pm 0.004~\mathrm{J~m}^{-2}$ for 100\% of relative humidity.
Contact adhesion measurements by \citet{wan-et-al1992} gave the surface energy $\gamma \approx 0.062~\mathrm{J~m}^{-2}$ for silica plates over a wide range of relative humidities (5--95\%).
\citet{derjaguin-et-al1977,derjaguin-et-al1978} applied an SFA to determine the Hamaker constant for fused quartz filaments, which can be translated to the surface energy $\gamma \approx 0.022~\mathrm{J~m}^{-2}$ within an accuracy of 10--20\% \citep[see][]{israelachvili2011}.

\subsection{Molecular Modeling}

The surface energy of amorphous silica in a vacuum could be determined by a molecular dynamics (MD) simulation, but it turned out that the value of the surface energy strongly depends on the interatomic potential adopted for the study.
\citet{roder-et-al2001} derived $\gamma \sim 0.900~\mathrm{J~m}^{-2}$ at $4427^\circ\mathrm{C}$\footnote{Note that MD simulations consider only nanoseconds, while the sublimation of silica requires tens of microseconds at a temperature of $4427^\circ\mathrm{C}$.} and $\gamma \sim 1.250~\mathrm{J~m}^{-2}$ at $2477^\circ\mathrm{C}$ from their MD simulation of amorphous silica nanoparticles with siloxane ({a functional group with the Si--O--Si linkage}) surfaces using the so-called BKS potential.
In contrast, \citet{hoang2007} obtained $\gamma \approx 0.1~\mathrm{J~m}^{-2}$ for siloxane surfaces at a temperature of $77^\circ\mathrm{C}$ and $\gamma \approx 0.142~\mathrm{J~m}^{-2}$ at $6727^\circ\mathrm{C}$ using a Morse-type potential for short-range interactions.
MD simulations of amorphous silica nanoparticles with siloxane surfaces also showed that the surface energy depends on the size and temperature of the particles if their sizes are as small as a few nanometers \citep{roder-et-al2001,hoang2007}.
MD simulations with the COMPASS force field by \citet{sun-et-al2013} have demonstrated the validity of the JKR theory to silica nanospheres.
They also derived the surface energy of $\gamma \approx 0.0295~\mathrm{J~m}^{-2}$ for siloxane surfaces from the maximum negative value of the Lennard-Jones potential.
Recently, \citet{leroch-wendland2013} have used a Morse-type potential for short-range interactions to simulate pull-off forces between silica nanospheres with a silanol density of $\alpha_\mathrm{OH} = 4.6~\mathrm{OH/nm^2}$ at room temperature under ambient conditions.
They found that their results are consistent with $\gamma \approx 0.085$--$0.091~\mathrm{J~m}^{-2}$ for relative humidities between 10\% and 80\%, and $\gamma = 0.034~\mathrm{J~m}^{-2}$ for dry air.
The surface energies for amorphous silica surfaces with $\alpha_\mathrm{OH} = 3.0~\mathrm{OH/nm^2}$ computed by \citet{leroch-wendland2012} with a Morse-type potential lie in the range of $\gamma=0.03$--$0.10~\mathrm{J~m}^{-2}$ at room temperature under ambient conditions depending on relative humidity.
\citet{cole-et-al2007} computed the surface energies of $\gamma=0.0365$--$0.097~\mathrm{J~m}^{-2}$ for oxidized silicon surfaces as well as $\gamma=0.019$--$0.090~\mathrm{J~m}^{-2}$ for amorphous silica surfaces at room temperature, based on the combination and extension of the Stillinger--Weber potential and the Vashishta potential.
\citet{cabriolu-ballone2010} obtained $\gamma = 0.31 \pm 0.02~\mathrm{J~m}^{-2}$ by using the Keating potential to model the siloxane surfaces of amorphous silica at room temperature.

\subsection{Contact Angle Techniques}

\citet{janczuk-zdziennicka1994} used a quartz glass cell to derive the surface energy of $\gamma = 0.058$--$0.060~\mathrm{J~m}^{-2}$ from contact angle measurements and suggested multiple layers of adsorbed water on the surface.
Contact angle measurements by \citet{zdziennicka-et-al2009} resulted in the surface energy of $\gamma = 0.059~\mathrm{J~m}^{-2}$ for quartz glass plates.
\citet{harnett-et-al2007} measured contact angles to calculate the surface energy of $\gamma = 0.0354 \pm 0.0035~\mathrm{J~m}^{-2}$ for original silicon dioxide chips and $\gamma = 0.0446 \pm 0.0031~\mathrm{J~m}^{-2}$ for cleaned ones with appropriate solvents.

\citet{kessaissia-et-al1981} derived the surface energy $\gamma = 0.151 \pm 0.007~\mathrm{J~m}^{-2}$ for mesoporous silica particles (Spherosil) with silanol density of $\alpha_\mathrm{OH} \approx 3.4 \pm 0.1~\mathrm{OH/nm^2}$ pressed into a disk from measurements of liquid contact angles and adsorption isotherms.
They also determined the surface energies for silica surfaces covered with hydrocarbon chains that were found to decrease with the length of the chains.
\citet{helmy-et-al2007} derived $\gamma = 0.2504~\mathrm{J~m}^{-2}$ for silica gel from isotherms and a different formula to calculate the surface energy with a monolayer of water molecules at room temperature.
They also reanalyzed the adsorption isotherms published in the literature inclusive of \citet{kessaissia-et-al1981} and obtained $\gamma = 0.1967$--$0.2826~\mathrm{J~m}^{-2}$ for silica samples and  $\gamma = 0.1728$--$0.2428~\mathrm{J~m}^{-2}$ for silicas outgassed at high temperatures.
\citet{{kimura-et-al2000}} determined the surface energy of $\gamma = 0.093~\mathrm{J~m}^{-2}$ for original silica filler using adsorption isotherms and $\gamma = 0.048$--$0.102~\mathrm{J~m}^{-2}$ for surface-modified silica filers.
When the original silica filler was heated at $200^\circ\mathrm{C}$ in He gas for 12~hr or at $400^\circ\mathrm{C}$ and $900^\circ\mathrm{C}$ in N$_2$ gas prior to measurements, the surface energies were found to be $\gamma = 0.1557$, 0.1181, and $0.0814~\mathrm{J~m}^{-2}$, respectively \citep{kimura-et-al2004}.
They performed the X-ray diffraction analysis and confirmed amorphous structures for the silica fillers heated to $200^\circ\mathrm{C}$ and $400^\circ\mathrm{C}$, but crystalline structures of cristobalite and tridymite for the silica filler heated to $900^\circ\mathrm{C}$.

\subsection{Thin-layer Wicking Technique}

Thin-layer wicking experiments carried out by \citet{chibowski-holysz1992} gave a surface energy of $\gamma = 0.0465$--$0.0603~\mathrm{J~m}^{-2}$ at room temperature for thin-layer chromatography silica, which were dried at $150^\circ\mathrm{C}$ for 2~hr prior to the experiments.
\citet{holysz1998} measured the wicking rates at room temperature to determine the surface energy components for thermally pretreated silica gel samples (Merck).
She obtained a total surface energy of $\gamma = 0.0488~\mathrm{J~m}^{-2}$ and a silanol density of $\alpha_\mathrm{OH} \approx 3.52~\mathrm{OH/nm^2}$ for $200^\circ\mathrm{C}$ pretreatment.
The surface energy increases with pretreatment temperature from $\gamma = 0.0547~\mathrm{J~m}^{-2}$ for $400^\circ\mathrm{C}$ to $\gamma = 0.0634~\mathrm{J~m}^{-2}$ for $1000^\circ\mathrm{C}$.
Her estimates of silanol densities indicate that hydrogen bonding on silica surfaces is electron-donor interaction rather than electron-acceptor interaction.
\citet{gonzalezmartin-et-al2001} derived the surface energy of $\gamma = 0.0491$--$0.0495~\mathrm{J~m}^{-2}$ at room temperature from their measurements with silica gel for thin-layer chromatography that was dried at $150^\circ\mathrm{C}$ for 1~hr before the experiments.
\citet{cui-et-al2005} obtained $\gamma = 0.0495~\mathrm{J~m}^{-2}$ for silica gel thin-layer chromatography plates that were dried at $150^\circ\mathrm{C}$ for 1~hr prior to wicking measurements.

\subsection{Calorimetric Method}

\citet{brunauer-et-al1956} determined the total surface energy of $\gamma=0.259 \pm 0.003~\mathrm{J~m}^{-2}$ for dehydroxylated amorphous silica in a vacuum by the heat-of-dissolution method.
However, \citet{tarasevich2007} claimed that the condition for dehydroxylated amorphous silica in \citet{brunauer-et-al1956} assumes the presence of hydroxyl groups on the surface of the silica.
By adsorption-calometric determination, \citet{tarasevich2007} obtained the surface energies of amorphous silica gel and macroporous amorphous silica (Silochrome) with silanol groups to be $\gamma =0.275~\mathrm{J~m}^{-2}$ and $0.200~\mathrm{J~m}^{-2}$, respectively.

\subsection{Cleavage Method}

The surface energy of brittle materials such as amorphous silica can be derived from the strain energy release rate for an equilibrium crack \citep{gillis-gilman1964,sridhar-et-al1997}.
\citet{lucas-et-al1995} derived $\gamma = 4.25$--$4.30~\mathrm{J~m}^{-2}$ from the strain energy release rate for vitreous silica glass that were annealed for three hours at $1050^\circ\mathrm{C}$.
\citet{michalske-fuller1985} measured the closure and reopening forces on cracks in vitreous silica glass, yielding a surface energy of $\gamma \approx 0.075~\mathrm{J~m}^{-2}$ for humidities above 20\% and $\gamma \approx 0.030 \pm 0.003~\mathrm{J~m}^{-2}$ for humidities close to 1\%.
\citet{wiederhorn-johnson1971} reported the surface energy of $\gamma \approx 2.2 \pm 0.1~\mathrm{J~m}^{-2}$ for silica glass that was heated at temperatures above $500^\circ\mathrm{C}$ and cooled to room temperature.
\citet{cocheteau-et-al2013} obtained $\gamma \approx 0.084~\mathrm{J~m}^{-2}$ for fused silica glass surfaces at room temperature and $\gamma \approx 0.20$--$0.21~\mathrm{J~m}^{-2}$ when annealed at $100^\circ\mathrm{C}$.
\citet{lawn-et-al1987} used fused silica microscope slides to observe crack motion and determined $\gamma = 0.1~\mathrm{J~m}^{-2}$ under ambient conditions (55\% relative humidity).
\citet{kalkowski-et-al2010,kalkowski-et-al2011,kalkowski-et-al2012} obtained the surface energy of $\gamma = 0.3 \pm 0.1~\mathrm{J~m}^{-2}$ for fused silica wafers bonded in a vacuum and heated to $200^\circ\mathrm{C}$--$250^\circ\mathrm{C}$ for several hours.

A study on the bonding of oxidized silicon wafers is useful to better understand the surface energy of amorphous silica because the surfaces of oxidized silicon wafers are occupied by amorphous silicon dioxide \citep{ploessl-kraeuter1999}.
Using the crack-opening method, \citet{maszara-et-al1988} determined the surface energy of oxidized silicon wafers with hydrophilic surfaces at room and elevated temperatures in an inert atmosphere.
The surface energy of $\gamma =0.0300$--$0.0425~\mathrm{J~m}^{-2}$ was determined at room temperature, implying that only a fraction of silanol groups is in contact.
They attributed the surface energy of $\gamma =0.050$--$0.075~\mathrm{J~m}^{-2}$ at temperatures of $100^\circ\mathrm{C}$--$200^\circ\mathrm{C}$ to hydrogen bonds between silanol groups.
The temporal variation in the surface energy was observed at temperatures below $600^\circ\mathrm{C}$, which we could fit by $\gamma =0.085~\mathrm{J~m}^{-2} (t/10~\mathrm{s})^{1/n}$ with $n=8 \pm 1$ at $300^\circ\mathrm{C}$--$400^\circ\mathrm{C}$ \citep[cf.][]{maszara-et-al1988}.
The temporal variation was interpreted by \citet{maszara-et-al1988} as being due to the hydrogen bonds between silanol groups being replaced by siloxane bonds due to the loss of water molecules.
Siloxane surfaces on the oxidized silicon wafers were established at $600^\circ\mathrm{C}$--$1100^\circ\mathrm{C}$, as they observed nearly the independence of the surface energy on time.
The surface energies in this range of temperatures are $\gamma = 0.160$--$0.195~\mathrm{J~m}^{-2}$ at $600^\circ\mathrm{C}$, $\gamma = 0.205$--$0.230~\mathrm{J~m}^{-2}$ at $800^\circ\mathrm{C}$, $\gamma = 0.295$--$0.440~\mathrm{J~m}^{-2}$ at $1000^\circ\mathrm{C}$, $\gamma = 0.34$--$0.55~\mathrm{J~m}^{-2}$ at $1100^\circ\mathrm{C}$, and $\gamma \approx 1.1~\mathrm{J~m}^{-2}$ at $1400^\circ\mathrm{C}$.
\citet{li-et-al2013} determined the surface energy of $\gamma = 0.05~\mathrm{J~m}^{-2}$ for oxidized silicon wafers at room temperature and $\gamma = 0.76$--$0.11~\mathrm{J~m}^{-2}$ after 2~hr of annealing at $200^\circ\mathrm{C}$--$350^\circ\mathrm{C}$ in N$_2$ atmosphere, which was accounted for by the transformation of hydrogen bonds into siloxane bridges.
They also reported that the surface energies at annealing temperatures of $500^\circ\mathrm{C}$--$1000^\circ\mathrm{C}$ reach $\gamma = 0.23$--$0.90~\mathrm{J~m}^{-2}$ and the surface energies increase when the surface was activated by O$_2$ plasma.
\citet{fournel-et-al2012} have shown that the surface energies for oxidized silicon wafers in an anhydrous nitrogen atmosphere were higher than those in an ambient atmosphere from $20^\circ\mathrm{C}$ to $1200^\circ\mathrm{C}$.
The measured surface energies were $\gamma \approx 0.071$--$0.10~\mathrm{J~m}^{-2}$ at $20^\circ\mathrm{C}$, $\gamma \approx 0.27$--$0.38~\mathrm{J~m}^{-2}$ at $200^\circ\mathrm{C}$, and $\gamma \approx 0.55$--$1.2~\mathrm{J~m}^{-2}$ at $400^\circ\mathrm{C}$.
At temperatures higher than $400^\circ\mathrm{C}$, the surface energies under an ambient atmosphere increase gradually to $\gamma \approx 1.1~\mathrm{J~m}^{-2}$ at $1100^\circ\mathrm{C}$ and steeply to $\gamma \approx 2.4~\mathrm{J~m}^{-2}$ at $1200^\circ\mathrm{C}$, but those in an anhydrous nitrogen atmosphere fluctuate in the range of $\gamma \approx 1.2$--$2.9~\mathrm{J~m}^{-2}$.
\citet{wang-suga2011} studied the influence of O$_2$ plasma and fluorine containing plasma on the surface energies for oxidized silicon wafers bonded in air.
They obtained $\gamma \approx 0.3~\mathrm{J~m}^{-2}$ for O$_2$ treatment, $\gamma \approx 1.2~\mathrm{J~m}^{-2}$ for O$_2 + $CF$_4$ treatment, and $\gamma = 0.046 \pm 0.015~\mathrm{J~m}^{-2}$ for no plasma treatment at room temperature.
\citet{turner-spearing2006} determined the surface energy of $\gamma =0.018$--$0.020~\mathrm{J~m}^{-2}$ for oxidized silicon wafers at room temperature under ambient conditions.
\citet{eichler-et-al2010} derived the surface energies for oxidized silicon wafers as well as fused silica wafers from in-situ crack length measurements under ambient conditions.
The surface energies for oxidized silicon and fused silica at room temperature were $\gamma \approx 0.046~\mathrm{J~m}^{-2}$ and $\gamma \approx 0.024~\mathrm{J~m}^{-2}$, respectively.
They have shown that the surface energies increase with wafer temperature and the values were converged to $\gamma \approx 0.205~\mathrm{J~m}^{-2}$ for oxidized silicon and $\gamma \approx 0.27~\mathrm{J~m}^{-2}$ for fused silica at $200^\circ\mathrm{C}$.
The surface energies for oxidized silicon wafers in air determined by \citet{tong-et-al1994} were $\gamma \approx 0.073 \pm 0.011~\mathrm{J~m}^{-2}$ at room temperature and increased with temperature up to $\gamma \approx 0.24 \pm 0.07~\mathrm{J~m}^{-2}$ at $150^\circ\mathrm{C}$.
\citet{suni-et-al2002} measured the surface energies of $\gamma \approx 0.50$--$0.75~\mathrm{J~m}^{-2}$ for oxidized silicon wafers in the temperature range of 100--$400^\circ\mathrm{C}$ activated in reactive oxygen plasma prior to bonding in air and then annealed for 2~hr at $100^\circ\mathrm{C}$.
\citet{pasquariello-et-al2000a,pasquariello-et-al2000b} measured the surface energy for oxidized silicon wafers bonded in oxygen plasma at room temperature using the razor-blade method in comparison to the surface energy of $\gamma \approx 0.033~\mathrm{J~m}^{-2}$ measured under ambient conditions.
The measurements were also performed at elevated temperatures, yielding $\gamma \approx 0.31~\mathrm{J~m}^{-2}$ at $480^\circ\mathrm{C}$ and $\gamma \approx 0.51~\mathrm{J~m}^{-2}$ at $720^\circ\mathrm{C}$ without exposure to oxygen plasma and the corresponding values with exposure to oxygen plasma were even higher \citep{pasquariello-et-al2000b}.
\citet{pasquariello-hjort2000} applied the mesa-spacer method to determine the surface energy of $\gamma \approx 0.020 \pm 0.002~\mathrm{J~m}^{-2}$ for oxidized silicon wafers at room temperature.
They have shown that the surface energy determined by the mesa-spacer method agrees with that determined by the razor-blade method after appropriate corrections for the infrared resolution limit.

\subsection{Optical Spectroscopy Techniques}

The surface energy of solids may be calculated within a 10--20\% accuracy from the Hamaker constant, which can be derived from the optical properties and electronic structure of the solids \citep{israelachvili2011}.
\citet{wittmann-et-al1971} used the full spectral method to determine the Hamaker constant for quartz glass at room temperature around a relative humidity of 1\%, corresponding to $\gamma \approx 0.030~\mathrm{J~m}^{-2}$.
The Hamaker constant for amorphous SiO$_2$ (Suprasil$^{\textregistered}$, Heraeus) calculated with the full spectral method by \citet{tan-et-al2003,tan-et-al2005} is equivalent to $\gamma \approx 0.035~\mathrm{J~m}^{-2}$.
\citet{french-et-al1995} calculated the Hamaker constants for fused silica glass (Suprasil$^{\textregistered}$, Heraeus), equivalent to $\gamma \approx 0.032~\mathrm{J~m}^{-2}$ and $\gamma \approx 0.034~\mathrm{J~m}^{-2}$ using the full spectral method and the Tabor--Winterton approximations, respectively \citep[see, also][]{ackler-et-al1996,french2000}.
\citet{hough-white1980} used the simple spectral method to calculate the Hamaker constant for fused silica, resulting in $\gamma \approx 0.032~\mathrm{J~m}^{-2}$ \citep[see also][]{bergstroem-et-al1996,bergstroem1997}.

\subsection{Ultrafast Opto-acoustic Technique}

\citet{ayouch-et-al2012} applied the ultrafast opto-acoustic technique to determine the surface energies for silica nanoparticles covered with either ethoxy groups or hydroxyl groups.
The surface energy of silica nanoparticles with hydroxyl groups was $\gamma = 0.028$--$0.050~\mathrm{J~m}^{-2}$ higher than $\gamma =  0.002$--$0.009~\mathrm{J~m}^{-2}$ for silica nanoparticles with ethoxy groups.

\subsection{Theoretical Consideration}

\citet{iler1979} estimated the surface energy of $\gamma = 0.275~\mathrm{J~m}^{-2}$ for vitreous silica from the surface tension of melted glass measured at $1300^\circ\mathrm{C}$ and extrapolated to zero alkali content and $25^\circ\mathrm{C}$.
\citet{parks1984} argued that the surface energy of silica glass should exceed $\gamma=0.275~\mathrm{J~m}^{-2}$ at room temperature by taking into account the entropy of the silica surface.
\citet{brunauer-et-al1956} derived the total surface energy of $\gamma=0.129 \pm 0.008~\mathrm{J~m}^{-2}$ for hydroxylated amorphous silica with silanol surfaces from their measured value for dehydroxylated amorphous silica and the heat of hydration, although the presence of adsorbed water molecules on the surface of the silica is implicitly assumed \citep{tarasevich2007}.
\citet{michalske-fuller1985} estimated the surface energy of $\gamma = 0.085~\mathrm{J~m}^{-2}$ for fully hydrated silica under ambient conditions from the number of possible hydrogen-bonding sites per unit area times the energy per hydrogen bond linkage.
\citet{stengl-et-al1989} considered the bonding process during annealing to determine the surface energies of $\gamma = 0.052$, $0.317$, and $0.98~\mathrm{J~m}^{-2}$ for oxidized silicon wafers bonded by H$_2$O molecules, silanol groups, and siloxane bridges, respectively.
A model on the dynamics of oxidized silicon wafer surfaces during annealing developed by \citet{han-et-al2000} gives a surface energy of $\gamma \approx 0.1~\mathrm{J~m}^{-2}$ for thermal oxides at room temperature.
They estimated the saturated surface energies of $\gamma =0.414~\mathrm{J~m}^{-2}$ at temperatures of $600^\circ\mathrm{C}$--$800^\circ\mathrm{C}$ and $\gamma =2.439~\mathrm{J~m}^{-2}$ at $800^\circ\mathrm{C}$--$1200^\circ\mathrm{C}$.
\citet{reiche2008} calculated the surface energy for oxidized hydrophilic silicon wafers to be $\gamma =0.165~\mathrm{J~m}^{-2}$ using hydrogen bond energies of isolated and associated silanol groups.
He claims that the surface energy increases to $\gamma =0.23$--$0.25~\mathrm{J~m}^{-2}$ when stored for a long period at room temperature.
Using a Morse-type potential energy, \citet{tromans-meech2004} estimated the surface energy of $\gamma = 1.8608~\mathrm{J~m}^{-2}$ for silica glass with covalently bonded siloxane surfaces at room temperature.
\citet{shchipalov2000} derived $\gamma =1.522~\mathrm{J~m}^{-2}$ for quartz glass from theoretical analyses of the structures and the experimentally determined zeta-potential.

\subsection{Synthetic Values}

\begin{figure}
\epsscale{.60}
\plotone{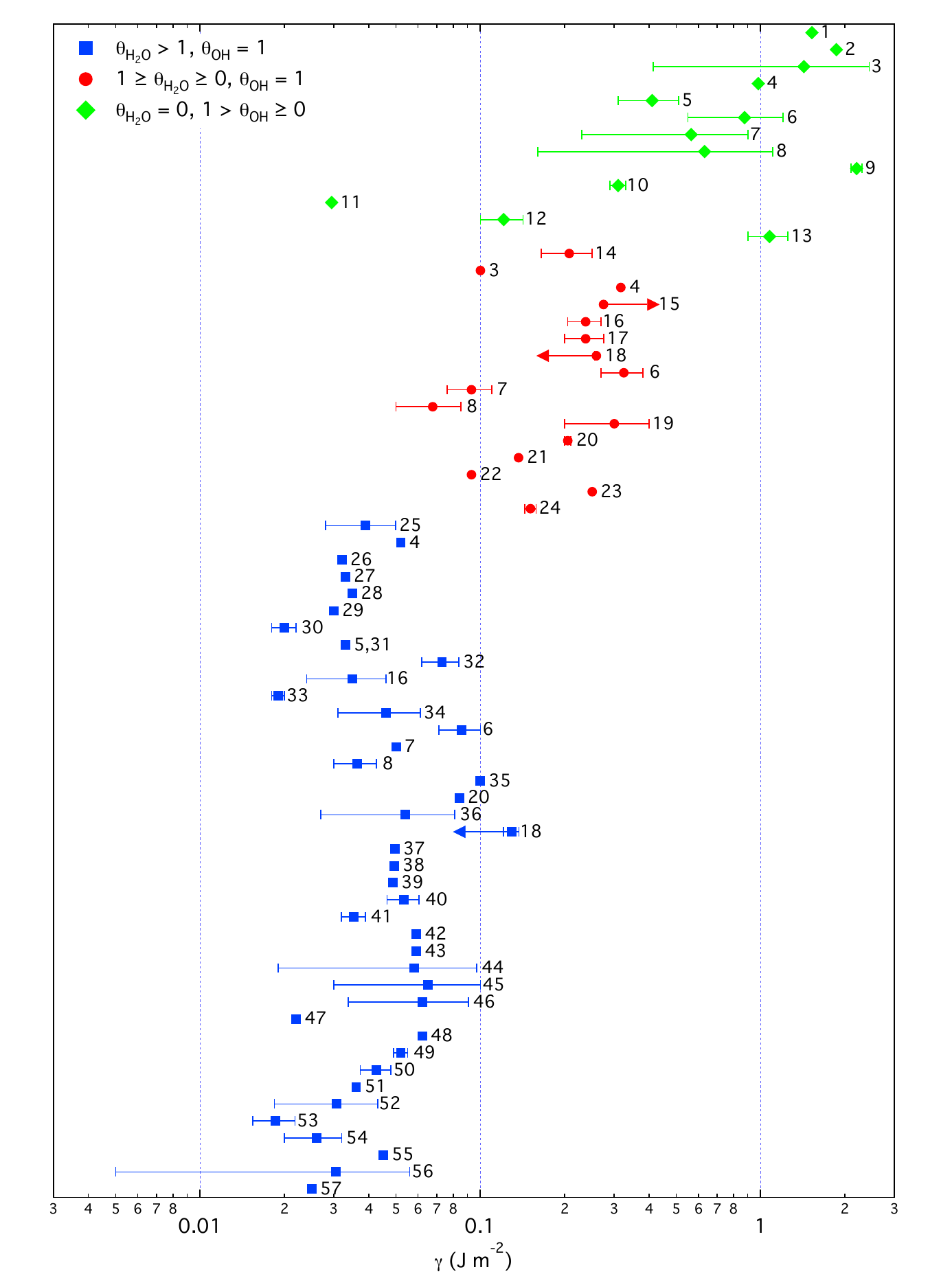}
\caption{Surface energy $\gamma$ for amorphous silica under different environments.
Filled squares: at room temperature in air; filled circles: at room temperature in a vacuum; filled diamonds: at elevated temperatures. 
The numbers beside the symbols indicate the data sources of the surface energy: (1) \citet{shchipalov2000}, (2) \citet{tromans-meech2004}, (3) \citet{han-et-al2000}, (4) \citet{stengl-et-al1989}, (5) \citet{pasquariello-et-al2000b}, (6) \citet{fournel-et-al2012}, (7) \citet{li-et-al2013}, (8) \citet{maszara-et-al1988}, (9) \citet{wiederhorn-johnson1971}, (10) \citet{cabriolu-ballone2010}, (11) \citet{sun-et-al2013}, (12) \citet{hoang2007}, (13) \citet{roder-et-al2001}, (14) \citet{reiche2008}, (15) \citet{parks1984}, (16) \citet{eichler-et-al2010}, (17) \citet{tarasevich2007}, (18) \citet{brunauer-et-al1956}, (19) \citet{kalkowski-et-al2010,kalkowski-et-al2011,kalkowski-et-al2012}, (20) \citet{cocheteau-et-al2013}, (21) \citet{kimura-et-al2004}, (22) \citet{kimura-et-al2000}, (23) \citet{helmy-et-al2007}, (24) \citet{kessaissia-et-al1981}, (25) \citet{ayouch-et-al2012}, (26) \citet{hough-white1980}, (27) \citet{french-et-al1995}, (28) \citet{tan-et-al2003,tan-et-al2005}, (29) \citet{wittmann-et-al1971}, (30) \citet{pasquariello-hjort2000}, (31) \citet{pasquariello-et-al2000a}, (32) \citet{tong-et-al1994}, (33) \citet{turner-spearing2006}, (34) \citet{wang-suga2011}, (35) \citet{lawn-et-al1987}, (36) \citet{michalske-fuller1985}, (37) \citet{cui-et-al2005}, (38) \citet{gonzalezmartin-et-al2001}, (39) \citet{holysz1998}, (40) \citet{chibowski-holysz1992}, (41) \citet{harnett-et-al2007}, (42) \citet{zdziennicka-et-al2009}, (43) \citet{janczuk-zdziennicka1994}, (44) \citet{cole-et-al2007}, (45) \citet{leroch-wendland2012}, (46) \citet{leroch-wendland2013}, (47) \citet{derjaguin-et-al1977,derjaguin-et-al1978}, (48) \citet{wan-et-al1992}, (49) \citet{horn-et-al1989}, (50) \citet{ling-et-al2007}, (51) \citet{ecke-butt2001}, (52) \citet{ecke-et-al2001}, (53) \citet{heim-et-al1999}, (54) \citet{fuji-et-al1999}, (55) \citet{bradley1932}, (56) \citet{vigil-et-al1994}, (57) \citet{kendall-et-al1987a,kendall-et-al1987b}.
\label{fig1}}
\end{figure}

We shall synthesize the data on the surface energy of amorphous silica with the surface chemistry, which depends on the environments.
Adsorbed H$_2$O molecules under ambient conditions can be removed from the surface of amorphous silica if the surface is heated or brought into a vacuum \citep{christy2010}.
As a result, the surface of amorphous silica at room temperature is covered by multilayers of H$_2$O molecules under ambient conditions and a monolayer of H$_2$O molecule in a vacuum \citep{vigil-et-al1994,zhuravlev2000}.
Moreover, the surface of amorphous silica in a vacuum attains a monolayer of H$_2$O molecules at temperatures of $25^\circ\mathrm{C}$--$190^\circ\mathrm{C}$, silanol groups at temperatures of $190^\circ\mathrm{C}$--$400^\circ\mathrm{C}$, and siloxane bridges develop at temperatures above $400^\circ\mathrm{C}$ \citep{zhuravlev2000}.
Hereafter, we divide the surface chemistry into three categories: (A) multilayers of H$_2$O molecules ($\theta_\mathrm{H_2O} > 1$, $\theta_\mathrm{OH} = 1$); (B) a monolayer of H$_2$O molecule and silanol groups ($1 \ge \theta_\mathrm{H_2O} \ge 0$, $\theta_\mathrm{OH} = 1$); and (C) siloxane groups ($\theta_\mathrm{H_2O} = 0$, $1 > \theta_\mathrm{OH} \ge 0$).
We do not include the surface energies measured after activation in a reactive ion, since the above-mentioned simple categorization cannot be applied to these activated surfaces and these conditions are not relevant to collision experiments.
Figure~\ref{fig1} is compiled from the data on the surface energies of amorphous silica surfaces classified into the above-mentioned categories by the surface chemistry. 
Here we consider that the reduced pressure represents environments where the surfaces of amorphous silica are free from multiple layers of water molecules ($\theta_\mathrm{H_2O} \le 1$), while the ambient conditions represent environments where the surfaces of amorphous silica are covered by multiple layers of adsorbed water molecules ($\theta_\mathrm{H_2O} > 1$).
Owing to large variations in experimentally and theoretically derived values for the surface energy of amorphous silica, we cannot place a tight constraint on the surface energy of amorphous silica in different environments.
Nevertheless, it may be safe to expect that the surface energy for amorphous silica surfaces lies in the range of $\gamma \la 0.1~\mathrm{J~m}^{-2}$ at room temperature under ambient conditions (A),  $\gamma=0.1$--$0.3~\mathrm{J~m}^{-2}$ at room temperature in reduced pressure (B), and $\gamma \ga 0.3~\mathrm{J~m}^{-2}$ above $400^\circ\mathrm{C}$ (C).
This picture is in accordance with the idea that the hydrogen bond between silanol groups is stronger than that between H$_2$O molecules, and the siloxane bond is the strongest \citep{stengl-et-al1989,fuji-et-al1999}.

\section{COMPARISONS OF THE JKR THEORY WITH COLLISION EXPERIMENTS}

\subsection{Critical Velocity for the Sticking of Silica Spheres to Silica Flat Surfaces}

\begin{figure}
\epsscale{.60}
\plotone{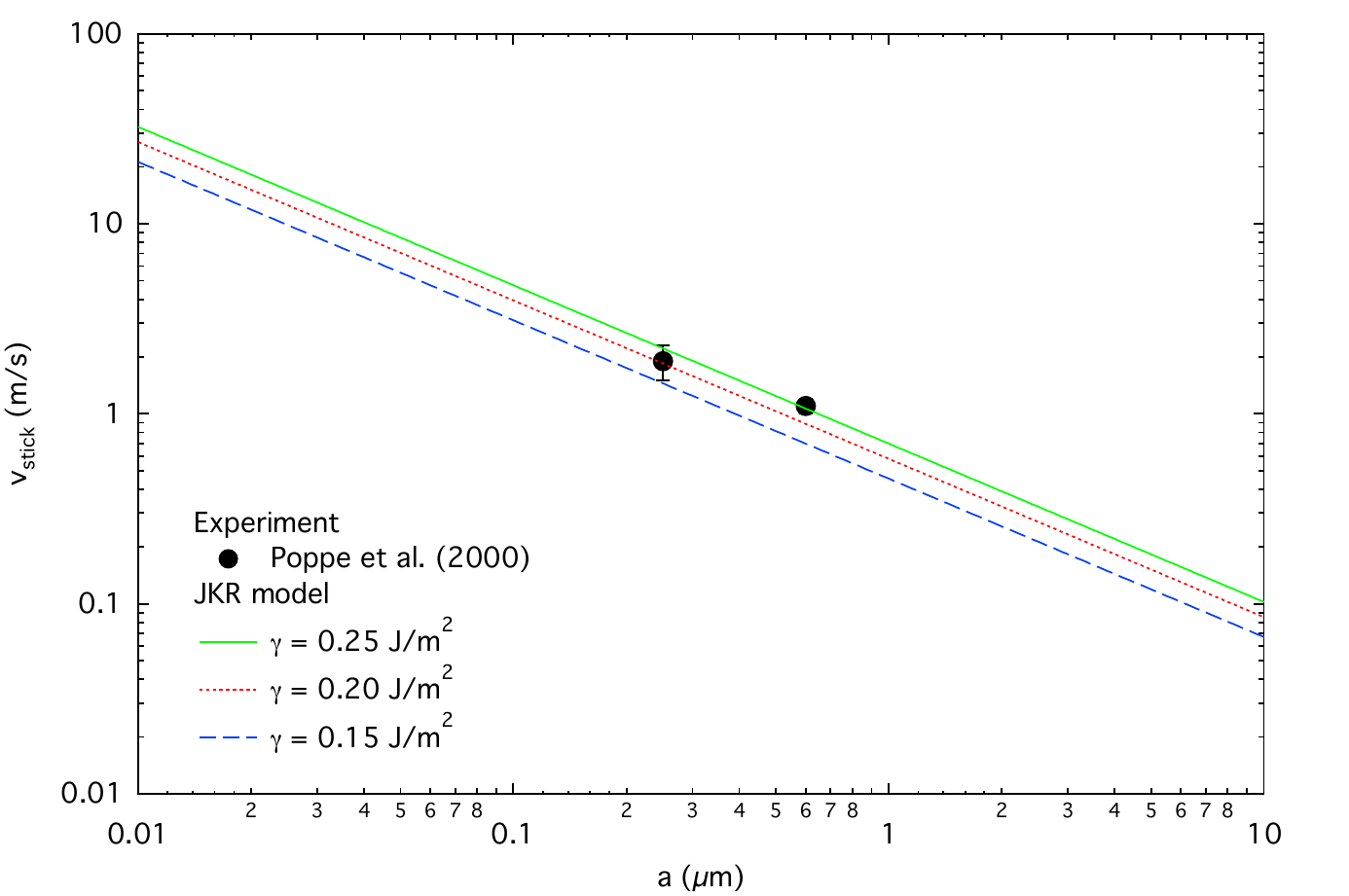}
\caption{Critical velocity for the sticking $v_\mathrm{stick}$ of amorphous silica spheres to an amorphous silica target as a function of sphere radius $a$. Solid line: Equation~(\ref{vstick}) with $\gamma \approx 0.25~\mathrm{J~m}^{-2}$; dotted line: Equation~(\ref{vstick}) with $\gamma \approx 0.20~\mathrm{J~m}^{-2}$; dashed line: Equation~(\ref{vstick}) with $\gamma \approx 0.15~\mathrm{J~m}^{-2}$. Filled circles: experimental values \citep{poppe-et-al2000}. \label{fig2}}
\end{figure}

\citet{poppe-et-al2000} experimentally determined the sticking probability of amorphous silica spheres to polished silica surfaces and oxidized silicon wafer surfaces in a vacuum ($1.4~\mathrm{Pa}$).
They observed that amorphous silica spheres with a radius of $a=0.6~\micron$ stick to the surfaces at velocities up to $v_\mathrm{stick} = 1.2 \pm 0.1~\mathrm{m~s^{-1}}$ and those with $a=0.25~\micron$ stick to the surfaces up to $v_\mathrm{stick} = 1.9 \pm 0.4~\mathrm{m~s^{-1}}$.
The critical velocity for sticking between a sphere and a flat surface is given by \citep{chokshi-et-al1993,dominik-tielens1997}
\begin{eqnarray}
v_\mathrm{stick}&=&{\left({\frac{27c_1\pi^{2/3}}{2^{5/3}}}\right)}^{1/2} {\left[{\frac{\gamma^5{\left({1-\nu^2}\right)}^2}{a^5\rho^3E^2}}\right]}^{1/6}, 
\label{vstick}
\end{eqnarray}
where $\rho$, $\nu$, and $E$ are the specific density, Poisson's ratio, and Young's modulus, respectively.
While the elastic properties of materials depend on the temperature, the deviations in the temperature range of interest are confined within 10\% of $E = 70~\mathrm{GPa}$ \citep[cf.][]{mcskimin1953,fine-et-al1954,spinner-cleek1960,rouxel2007}.
There is uncertainty in the value for $c_1$, which slightly varies from one definition of the threshold velocity to another, but $c_1 \approx 1$ \citep{chokshi-et-al1993,thornton-ning1998,brilliantov-et-al2007}. 
Figure~\ref{fig2} compares the critical velocities for sticking between an amorphous silica sphere and a flat surface expected by Equation~(\ref{vstick}) with the data obtained in the collision experiment by \citet{poppe-et-al2000}.
While we have constrained the synthetic value of the surface energy for amorphous silica in a vacuum to lie in the range of $\gamma=0.1$--$0.3~\mathrm{J~m}^{-2}$, the comparison with the collision experiments suggests $\gamma \approx 0.25~\mathrm{J~m}^{-2}$ as the most favorable value.
Therefore, we may write 
\begin{eqnarray}
v_\mathrm{stick}
&\approx&1.1~\mathrm{m~s^{-1}} {\left({\frac{c_1}{0.935}}\right)}^{1/2} \left(\left\{\frac{1}{0.9711}\right\}-\left\{\frac{0.0289}{0.9711}\right\}\left\{\frac{\nu}{0.17}\right\}^{2}\right)^{1/3} \left(\frac{a}{0.6~\micron}\right)^{-5/6} \nonumber \\
&& \times {\left(\frac{\gamma}{0.25~\mathrm{J~m^{-2}}}\right)}^{5/6} {\left(\frac{E}{70~\mathrm{GPa}}\right)}^{-1/3} \left(\frac{\rho}{2.0\times{10}^{3}~\mathrm{kg~m^{-3}}}\right)^{-1/2},
\label{vstick-value}
\end{eqnarray}
for collisions between amorphous silica spheres of $a=0.6~\micron$ and an amorphous silica surface.

\subsection{Critical Velocity for the Onset of Losing a Single Silica Sphere from a Silica Aggregate by Mutual Collision}

\begin{figure}
\epsscale{.60}
\plotone{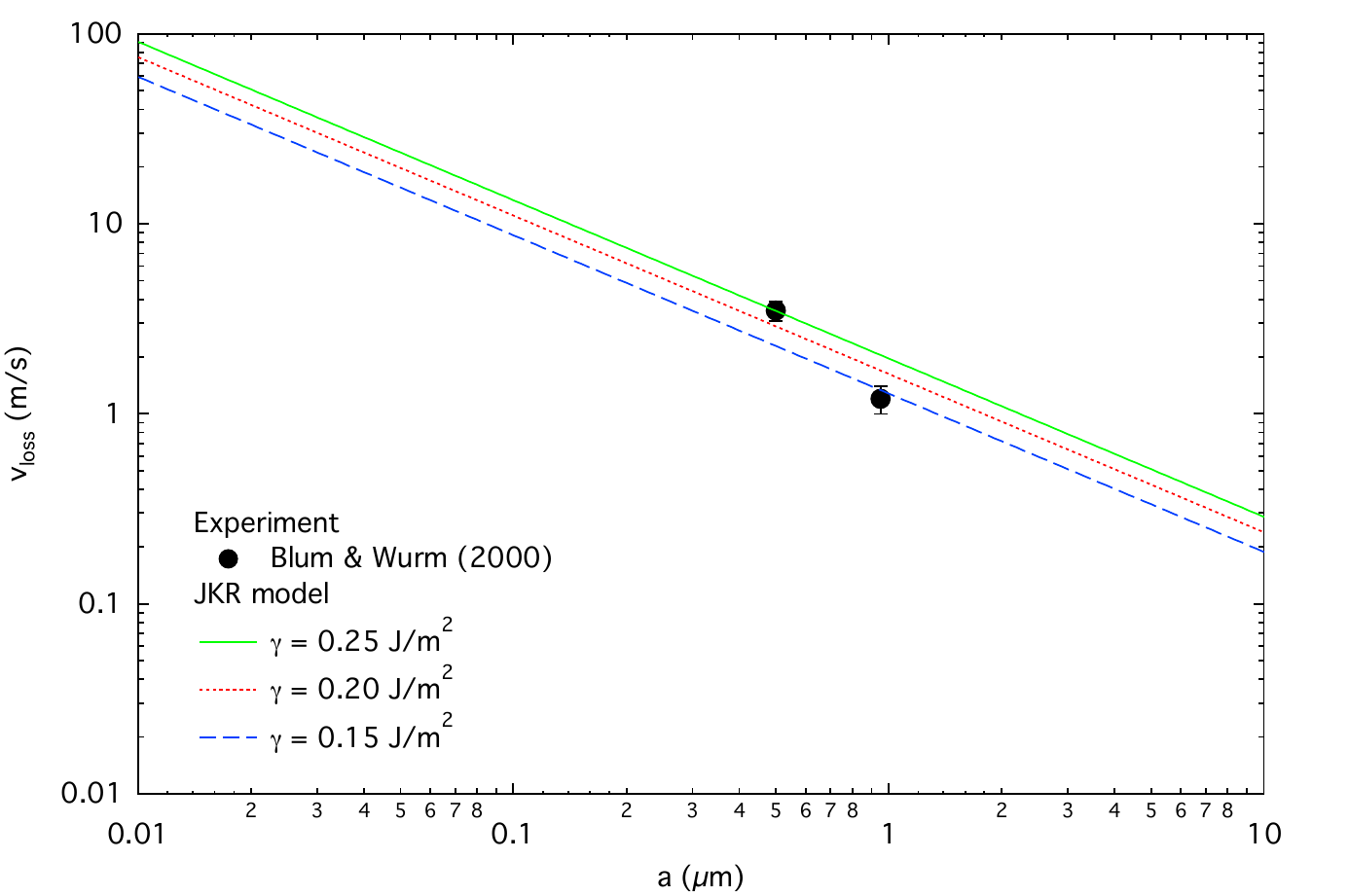}
\caption{Critical velocity for losing amorphous silica spheres from an aggregate consisting of the same spheres. Solid line: Equation~(\ref{vloss}) with $\gamma \approx 0.25~\mathrm{J~m}^{-2}$; dotted line: Equation~(\ref{vloss}) with $\gamma \approx 0.20~\mathrm{J~m}^{-2}$; dashed line: Equation~(\ref{vloss}) with $\gamma \approx 0.15~\mathrm{J~m}^{-2}$. Filled circles: experimental values \citep{blum-wurm2000}. \label{fig3}}
\end{figure}

\citet{blum-wurm2000} performed microgravity experiments on collision between aggregates of amorphous silica spheres to study their sticking, restructuring, and fragmentation in vacuum ($200~\mathrm{Pa}$).
They determined the onset of fragmentation at a velocity of $v_\mathrm{loss} = 1.2\pm0.2~\mathrm{m~s^{-1}}$ for $a=0.95~\micron$ and $v_\mathrm{loss} = 3.5\pm0.4~\mathrm{m~s^{-1}}$ for $a=0.50~\micron$.
In the framework of the JKR theory, the onset of fragmentation is located at a velocity when the kinetic energy per contact is three times the energy to break a contact \citep{dominik-tielens1997,wada-et-al2007}.
Therefore, the critical velocity for the onset of losing single spheres from an aggregate can be expressed as
\begin{eqnarray}
v_\mathrm{loss} &=& \sqrt{6} \, v_\mathrm{break},
\label{vloss}
\end{eqnarray}
where $v_\mathrm{break}$ is the velocity necessary to completely break a contact in the equilibrium position given by
\begin{eqnarray}
v_\mathrm{break} &=& \left({1.54\, \frac{27 \pi^{2/3}}{4}}\right)^{1/2} {\left[{\frac{\gamma^5{\left({1-\nu^2}\right)}^2}{a^5\rho^3E^2}}\right]}^{1/6} .
\label{vbreak}
\end{eqnarray}
Figure~\ref{fig3} shows a comparison of the critical velocities for the onset of losing single amorphous silica spheres from an aggregate derived from Equation~(\ref{vloss}) with those from the collision experiments by \citet{blum-wurm2000}.
It is worth noting that large silica particles of $a = 0.95~\micron$ were coated with dimethyldimethoxysilane (DMDMS), while small silica particles of $a = 0.50~\micron$ were uncoated.
We find that $\gamma \approx 0.25~\mathrm{J~m}^{-2}$ is consistent with the experimental results for $a = 0.50~\micron$, while $\gamma \approx 0.15~\mathrm{J~m}^{-2}$ fits better with the results for $a = 0.95~\micron$.
Consequently, we may write 
\begin{eqnarray}
v_\mathrm{loss} &=& 1.3~\mathrm{m~s^{-1}} \left(\left\{\frac{1}{0.9711}\right\}-\left\{\frac{0.0289}{0.9711}\right\}\left\{\frac{\nu}{0.17}\right\}^{2}\right)^{1/3} \left(\frac{a}{0.95~\micron}\right)^{-5/6} \nonumber \\
&& \times {\left(\frac{\gamma}{0.15~\mathrm{J~m^{-2}}}\right)}^{5/6} {\left(\frac{E}{70~\mathrm{GPa}}\right)}^{-1/3} \left(\frac{\rho}{2.0\times{10}^{3}~\mathrm{kg~m^{-3}}}\right)^{-1/2} ,
\label{vloss-value}
\end{eqnarray}
for collisions between aggregates consisting of DMDMS-coated amorphous silica spheres of $a=0.95~\micron$.

\subsection{Critical Velocity for the Restructuring of a Silica Aggregate by Mutual Collision}

\begin{figure}
\epsscale{.60}
\plotone{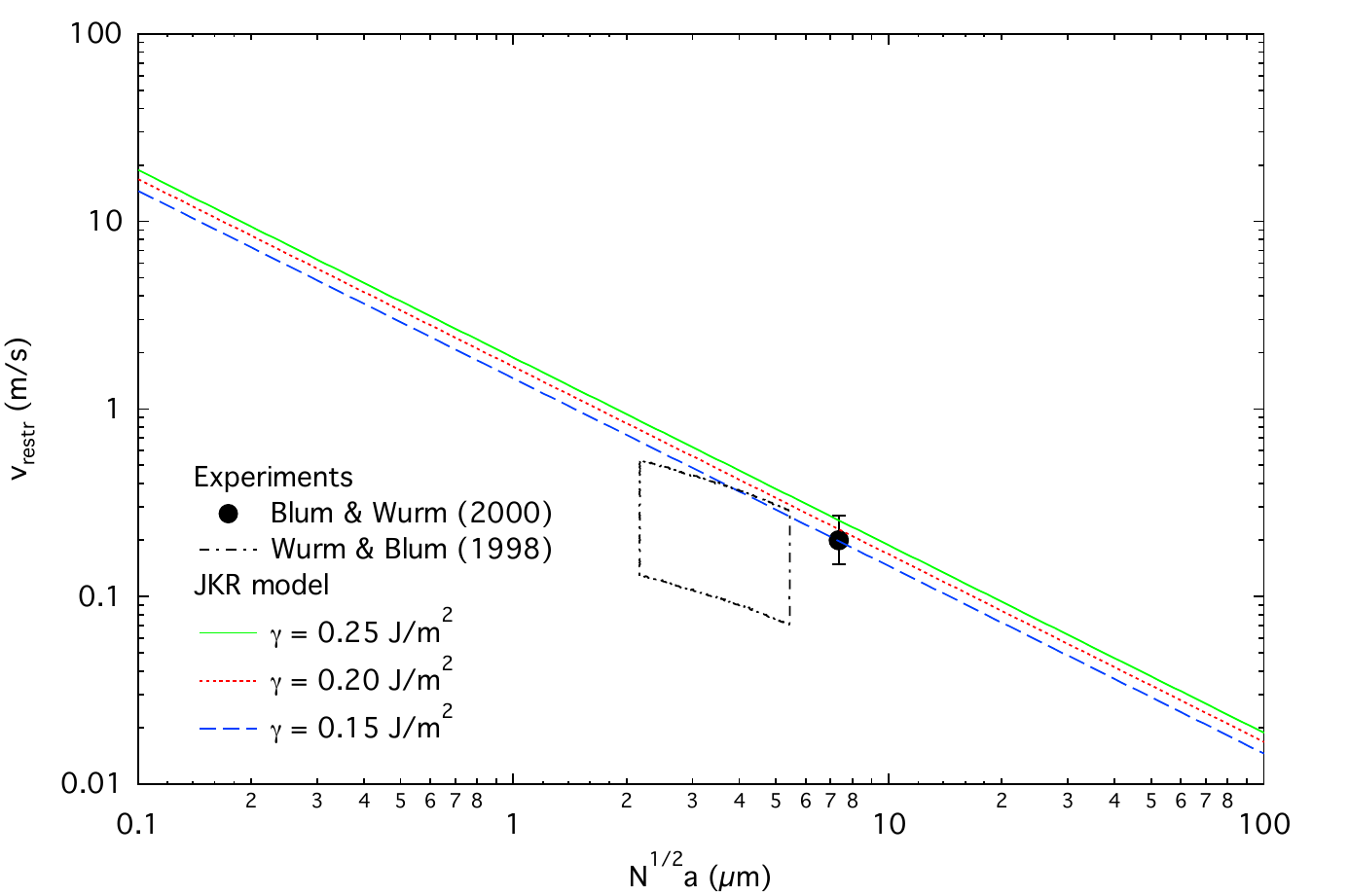}
\caption{Critical velocity for restructuring an aggregate consisting of amorphous silica spheres. Solid line: Equation~(\ref{vrestr}) with $\gamma \approx 0.25~\mathrm{J~m}^{-2}$; dotted line: Equation~(\ref{vrestr}) with $\gamma \approx 0.20~\mathrm{J~m}^{-2}$; dashed line: Equation~(\ref{vrestr}) with $\gamma \approx 0.15~\mathrm{J~m}^{-2}$. Filled circle and dashed-dotted line: experimental values \citep{wurm-blum1998,blum-wurm2000}. \label{fig4}}
\end{figure}

\citet{blum-wurm2000} determined the critical velocity $v_\mathrm{restr}$ for the onset of restructuring of aggregates consisting of DMDMS-coated amorphous silica spheres in a vacuum ($\sim 200~\mathrm{Pa}$)
The critical velocity $v_\mathrm{restr}$ is given by
\citep{dominik-tielens1997}
\begin{eqnarray}
v_\mathrm{restr}&=& \sqrt{\frac{30 E_\mathrm{roll}}{4 \pi \rho a^3 N}} , 
\label{vrestr}
\end{eqnarray}
where $N$ and $E_\mathrm{roll}$ are the number of spheres and the friction energy for the rolling of spheres by $90^\circ$, respectively.
The latter is given by
\begin{eqnarray}
E_\mathrm{roll}&=&6 \pi^2 \gamma a \xi_\mathrm{crit} , 
\end{eqnarray}
where $\xi_\mathrm{crit}$ is the critical displacement, above which the contact area between the particles starts to move, and is typically $\xi_\mathrm{crit} = 0.2~\mathrm{nm}$ \citep{dominik-tielens1995}.
The collision experiments resulted in $v_\mathrm{restr} = 0.20^{+0.07}_{-0.05}~\mathrm{m~s^{-1}}$ for aggregates consisting of 60 spheres of $a = 0.95~\micron$ \citep{blum-wurm2000}.
\citet{wurm-blum1998} used the same DMDMS-coated amorphous silica spheres to form aggregates in a turbomolecular pump using a ballistic cluster-cluster aggregation process without restructuring.
As a result, the range of collision velocities in their experiments is also able to constrain the critical velocity for the onset of the restructuring of aggregates in the range of $N = 5$--$33$.
Figure~\ref{fig4} demonstrates that Equation~(\ref{vrestr}) with $\gamma \approx 0.15~\mathrm{J~m}^{-2}$ is consistent with the experimental results of \citet{blum-wurm2000} and \citet{wurm-blum1998}.
We may therefore write
\begin{eqnarray}
v_\mathrm{restr}&=& 0.20~\mathrm{m~s^{-1}} \left({\frac{\gamma}{0.15~\mathrm{J~m^{-2}}}}\right)^{1/2} \left({\frac{a}{0.95~\micron}}\right)^{-1} \left({\frac{\rho}{2.0\times{10}^{3}~\mathrm{kg~m^{-3}}}}\right)^{-1/2} \nonumber \\
&& \times \left({\frac{\xi_\mathrm{crit}}{0.2~\mathrm{nm}}}\right)^{1/2} \left({\frac{N}{60}}\right)^{-1/2} , 
\label{vrestr-value}
\end{eqnarray}
with
\begin{eqnarray}
E_\mathrm{roll}&=&1.7 \times {10}^{-15}~\mathrm{J} \, \left({\frac{\gamma}{0.15~\mathrm{J~m}^{-2}}}\right) \left({\frac{a}{0.95~\micron}}\right) \left({\frac{\xi_\mathrm{crit}}{0.2~\mathrm{nm}}}\right) , 
\end{eqnarray}
for collisions between aggregates consisting of DMDMS-coated amorphous silica spheres of $a=0.95~\micron$.

\subsection{Erosion Rates of Aggregates Consisting of Amorphous Silica Spheres}

\begin{figure}
\epsscale{.60}
\plotone{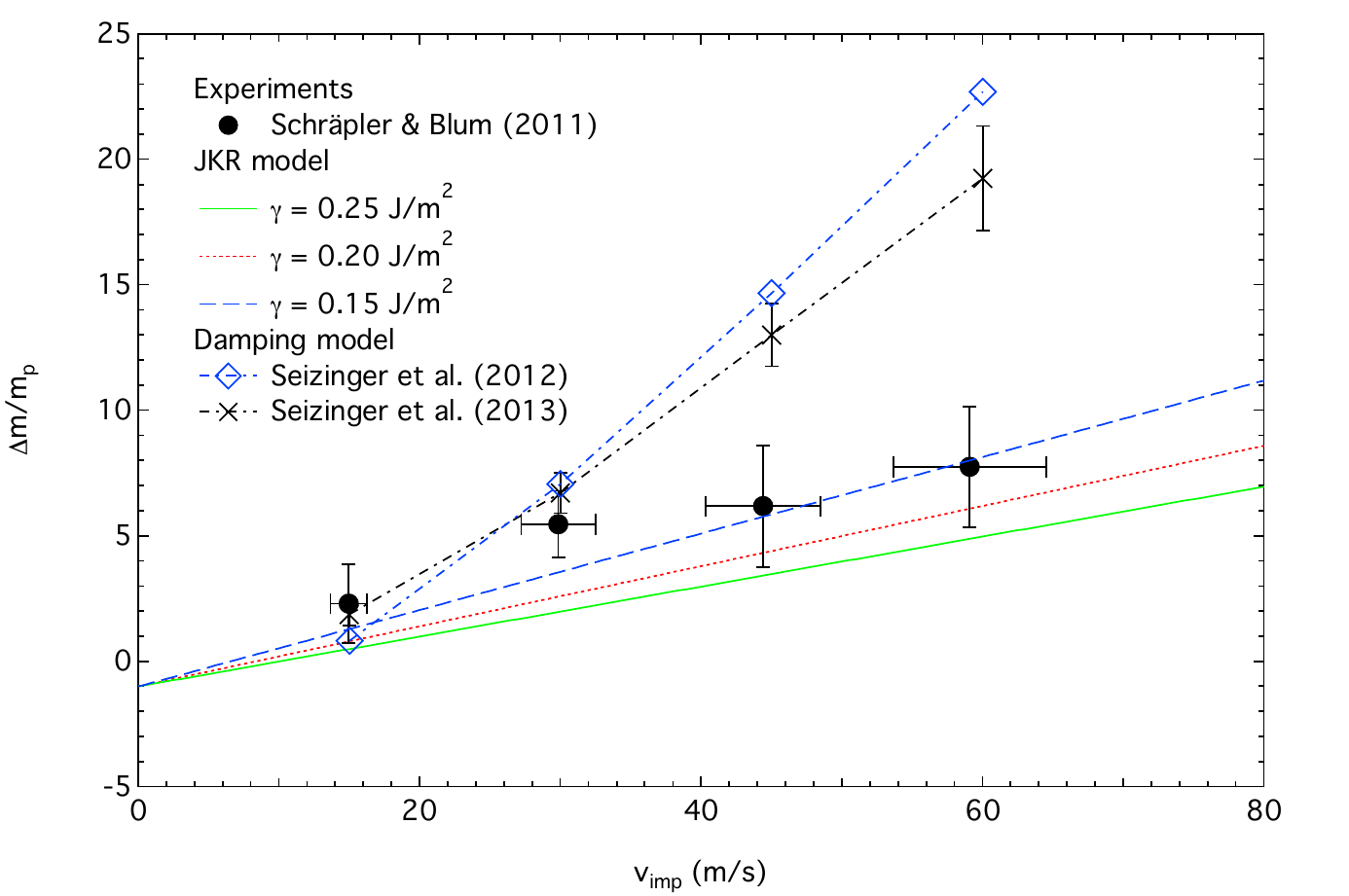}
\caption{Erosion efficiency of amorphous silica particles with $a=0.76~\micron$ colliding with aggregates consisting of the same silica particles. Solid line: Equation~(\ref{erosion-rate}) with $\gamma \approx 0.25~\mathrm{J~m}^{-2}$; dotted line: Equation~(\ref{erosion-rate}) with $\gamma \approx 0.20~\mathrm{J~m}^{-2}$; dashed line: Equation~(\ref{erosion-rate}) with $\gamma \approx 0.15~\mathrm{J~m}^{-2}$. Filled circles: experimental values \citep{schraepler-blum2011}. Crosses and open diamonds: damping models \citep{seizinger-et-al2012,seizinger-et-al2013}. \label{fig5}}
\end{figure}

\citet{schraepler-blum2011} defined the erosion rate of a target by the ratio of the change in the mass of the target before and after the exposure to dust impact, $\Delta m$, to the total mass of the projectiles, $m_\mathrm{p}$.
They derived experimentally the erosion rates for amorphous silica aggregates with a filling factor of $\phi \approx 0.15$ and $a = 0.76~\micron$ in a vacuum ($<4~\mathrm{Pa}$) as a function of impact velocity $v_\mathrm{imp}$ \citep[see][]{blum-schrapler2004}.
According to \citet{wada-et-al2013}, we can estimate the erosion rates $\Delta m/m_\mathrm{p}$ for aggregates by 
\begin{eqnarray}
\frac{\Delta m}{m_\mathrm{p}} &\approx& \frac{v_\mathrm{imp}}{\, v_\mathrm{disrupt}} -1 ,
\label{erosion-rate}
\end{eqnarray}
with
\begin{eqnarray}
v_\mathrm{disrupt} &=& {\epsilon}\, v_\mathrm{break},
\end{eqnarray}
where $v_\mathrm{disrupt}$ is the critical collision velocity, above which the growth of aggregates cannot be facilitated by coagulation.
Here $\epsilon$ is a non-dimensional coefficient that depends on the filling factor of the aggregates as well as the mass ratio of two colliding aggregates.
Numerical simulations on mutual collisions between aggregates suggest $\epsilon \simeq 10$ for high-mass ratios and $\epsilon \simeq 7.7$ for equal mass with a filling factor of $\phi \approx 0.15$, and $\epsilon \simeq 5.2$ for equal mass with $\phi \approx 0.01$ \citep{wada-et-al2008,wada-et-al2009,wada-et-al2013}.
Therefore, we may write
\begin{eqnarray}
\frac{\Delta m}{m_\mathrm{p}} + 1 &\approx& 5.0 \, \left({\frac{\epsilon}{10}}\right)^{-1} \left({\frac{v_\mathrm{imp}}{50~\mathrm{m/s}}}\right)\left({\frac{\gamma}{0.25~\mathrm{J~m}^{-2}}}\right)^{-5/6} \left(\left\{\frac{1}{0.9711}\right\}-\left\{\frac{0.0289}{0.9711}\right\}\left\{\frac{\nu}{0.17}\right\}^{2}\right)^{-1/3} \nonumber \\
&&\times \left({\frac{a}{0.76~\micron}}\right)^{5/6} \left(\frac{\rho}{2.0\times{10}^{3}~\mathrm{kg~m^{-3}}}\right)^{1/2} {\left(\frac{E}{70~\mathrm{GPa}}\right)}^{1/3} .
\label{e-schraepler}
\end{eqnarray}
for collisions between aggregates consisting of amorphous silica spheres of $a=0.76~\micron$.
Figure~\ref{fig5} compares the erosion rate for amorphous silica expected by Equation~(\ref{erosion-rate}) with the data obtained by the collision experiment with aggregates of $\phi \approx 0.15$ by \citet{schraepler-blum2011}.
Also plotted are the predictions by \citet{seizinger-et-al2013} who introduced a visco-elastic damping force in the JKR theory to fit the experimentally derived values of $\Delta m/m_\mathrm{p}$ \citep[see, also][]{seizinger-et-al2012}.
The prediction curves by \citet{seizinger-et-al2013} significantly overestimate the experimentally determined erosion rates at collision velocities of $v_\mathrm{imp} > 30~\mathrm{m~s^{-1}}$.
In contrast, Equation~(\ref{e-schraepler}) is consistent with the collision experiments by \citet{schraepler-blum2011}, although the experimental erosion rate at $v_\mathrm{imp} \approx 30~\mathrm{m~s^{-1}}$ is slightly higher than our prediction.
It is clear that Equation~(\ref{erosion-rate}) with $\gamma \approx 0.15~\mathrm{J~m}^{-2}$ shows a better agreement with the experiments compared with higher surface energies such as $\gamma \approx 0.25~\mathrm{J~m}^{-2}$.

\section{DISCUSSION}

\subsection{Controversial Experiments on the Effect of Circumstances}

We have shown that the well-known dependence of surface energy on circumstances has been the missing piece for harmonizing the collision experiments and the measurements of surface energy.
However, we are aware that \citet{bradley1932} and \citet{heim-et-al1999} claimed the independence of surface energy on the circumstances as opposed to the commonly accepted picture of the relation between the surface energy and the ambient pressure.
\citet{bradley1932} briefly stated that his results on the pull-off forces between silica spheres remained the same after the lower part of his apparatus was evacuated by means of a mercury vapor diffusion pump.
We speculate that his statement arose from either surface contamination by adsorption of mercury vapor or the condition that the spheres came into contact in air prior to evacuation.
\citet{heim-et-al1999} briefly mentioned that the surface energy did not change in the range of ambient pressure from $10^2$ to $10^5~\mathrm{Pa}$.
In contrast, \citet{ploessl-kraeuter1999} have shown that low vacuum of a few hundred pascals is sufficient to elevate the surface energy for oxidized silicon wafers by the removal of adsorbed H$_2$O molecules.
It is worth noting that the results in \citet{heim-et-al1999} differ from those in \citet{ecke-et-al2001}, \citet{ecke-butt2001}, and \citet{ling-et-al2007} despite of the same microspheres on their cantilevers.
While \citet{ecke-et-al2001} and \citet{ecke-butt2001} performed their measurements with oxidized silicon wafers, \citet{heim-et-al1999} and \citet{ling-et-al2007} used silica microspheres glued to a microscopy slide by epoxy heat resin.
It is well-known that the use of adhesive influences the determination of surface energy due to its surfactant contamination on the surface of the particle \citep{butt-et-al2005,castellanos2005,mak-et-al2006}.
\citet{heim-et-al1999} observed an extremely large scatter of adhesive forces for different pairs of particles, which could be accounted for by offset contacts between silica microspheres \citep[cf.][]{heim-et-al2004,yang-et-al2008}.
As noted by \citet{ling-et-al2007}, there is even a chance of contact between the epoxy and the colloidal probe during positioning, which transfers the epoxy to the surfaces of silica microspheres on the probe and the microscopy slide.
To clarify these speculations, we urge experimentalists to perform new measurements of surface energy for amorphous silica as a function of ambient pressure.

\subsection{Effect of DMDMS Coating on the Surface Energy of Amorphous Silica}

Aside from these conjectural issues, we have shown that collision experiments with amorphous silica are consistent with numerical models in the framework of the JKR theory, if accompanied by the surface energy for hydrophilic amorphous silica expected in a vacuum.
\citet{poppe-et-al2000} stated that the critical velocity for the sticking of amorphous silica spheres onto amorphous silica surfaces does not change significantly even if these surfaces are coated with a layer of DMDMS.
At first glance, this is at odds with the fact that the molecular structure of the outermost layer determines the surface energies of amorphous silica spheres.
\citet{poppe-et-al2000} claimed that the DMDMS-coated silica surfaces are hydrophobic surfaces whose surface energies should be much lower than those for hydrophilic ones \citep[see][]{fuji-et-al1999,ecke-butt2001,ecke-et-al2001,kappl-butt2002}.
On the contrary, \citet{blum-wurm2000} stated that the adhesion forces between the DMDMS-coated silica spheres were a factor of $1.35 \pm 0.23$ stronger than those measured by \citet{heim-et-al1999}.
This implies that the DMDMS density on the surface of amorphous silica surfaces used in \citet{poppe-et-al2000} and \citet{blum-wurm2000} was too low to be hydrophobic \citep[see][]{fuji-et-al1999}.
Moreover, Figure~\ref{fig3} shows that the DMDMS-coated silica spheres of $a=0.95~\micron$ used in \citet{blum-wurm2000} seem to have only slightly lower surface energies than the uncoated amorphous silica spheres of $a=0.50~\micron$.
Consequently, the collision experiments with the DMDMS-coated silica spheres performed by \citet{poppe-et-al2000} and \citet{blum-wurm2000} are also described by the model with hydrophilic amorphous silica in the framework of the JKR theory.

\subsection{Effect of Siloxane Density on the Surface Energy of Amorphous Silica}

As opposed to the optimal surface energy of $\gamma \approx 0.25~\mathrm{J~m}^{-2}$, a lower surface energy of $\gamma \approx 0.15~\mathrm{J~m}^{-2}$ seems to be more appropriate for the experimentally determined values of erosion rates in \citet{schraepler-blum2011} (see Figure~\ref{fig5}).
The former and the latter are consistent with the uncoated amorphous silica spheres used in both \citet{poppe-et-al2000} and \citet{blum-wurm2000}, and the DMDMS-coated amorphous silica spheres used in \citet{blum-wurm2000}, respectively.
The low surface energy cannot always be associated with surface coating, as \citet{schraepler-blum2011} did not describe any modifications to the surface of amorphous silica spheres \citep[see also][]{blum-schrapler2004}.
However, the surface chemistry of sicastar$^{\textregistered}$ used in \citet{schraepler-blum2011} may remarkably differ from that of amorphous silica spheres used in \citet{poppe-et-al2000} and \citet{blum-wurm2000}.
\citet{romeis-et-al2012} measured the Young's modulus of $E=44.7 \pm 7.6~\mathrm{GPa}$ for sicastar$^{\textregistered}$ and claimed a significantly reduced amount of siloxane bonds.
The relationship between the surface energy and the Young's modulus indicates that materials with a low value of Young's modulus have a low value of surface energy \citep{linford-mitchell1971,weir2008}.
Therefore, the low surface energy of $\gamma \approx 0.15~\mathrm{J~m}^{-2}$ for sicastar$^{\textregistered}$ used in the collision experiments by \citet{schraepler-blum2011} is a natural consequence of the low Young's modulus.
Consequently, the slight deviation of the model erosion rate with $\gamma \approx 0.25~\mathrm{J~m}^{-2}$ from experimental results does not violate the validity of the JKR theory.

\subsection{Size Dependence of the Critical Velocity for Sticking}

According to their results on collision experiments of amorphous silica spheres with $a = 0.25$ and $0.6~\micron$, \citet{poppe-et-al2000} proposed a power-law relation of $v_\mathrm{stick} \propto a^{-0.53}$ whose slope is gentler than $v_\mathrm{stick} \propto a^{-5/6}$ expected by the JKR theory.
In contrast, compared to that expected from the JKR theory, the collision experiments by \citet{blum-wurm2000} showed a steeper slope for the size dependence of $v_\mathrm{stick}$.
One might speculate that the the JKR theory cannot be applied to collision experiments since the theory is not intended to model dynamic contacts, but static or quasi-static contacts.
Nevertheless, the outcome of collision experiments has been shown to qualitatively agree with numerical simulations based on the JKR theory, if the empirical values for the break-up energy and the rolling friction force were used \citep{blum-wurm2000}.
As shown in Figure~\ref{fig2}, one may find that the theoretical prediction of $v_\mathrm{stick} \propto a^{-5/6}$ with $\gamma \approx 0.25~\mathrm{J~m}^{-2}$ fits the data by \citet{poppe-et-al2000} within the experimental error bars.
In the experiments by \citet{blum-wurm2000}, the theoretical prediction of $v_\mathrm{stick} \propto a^{-5/6}$ with $\gamma \approx 0.25~\mathrm{J~m}^{-2}$ fits the data with $a = 0.50~\micron$, but a slightly lower surface energy of $\gamma \approx 0.15~\mathrm{J~m}^{-2}$ is appropriate to the data with $a = 0.95~\micron$.
We could simply attribute the deviation from $v_\mathrm{stick} \propto a^{-5/6}$ in \citet{blum-wurm2000} to the slight difference in the surface treatment of amorphous silica spheres at different sizes.
Therefore, we conclude that they would have observed the $v_\mathrm{stick} \propto a^{-5/6}$ expected by the JKR theory, provided that amorphous silica spheres at different radii were identical in surface chemistry.

\subsection{Rolling Motion of Amorphous Silica Spheres}

The surface energy is a crucial parameter for controlling not only the normal motion, but also the rolling motion of elastic particles as formulated in Eqs.~(\ref{vstick-value}), (\ref{vloss-value}), and (\ref{vrestr-value}).
We have shown that the results from the collision experiments with the DMDMS-coated amorphous silica spheres in \citet{blum-wurm2000} are consistent with $\gamma \approx 0.15~\mathrm{J~m}^{-2}$ for both the motions (see Figures~\ref{fig3} and \ref{fig4}).
Their microscopic analyses of the collision experiments revealed a rolling friction force of $F_\mathrm{roll} = 0.50 \pm 0.25~\mathrm{nN}$ for the gravitational restructuring of the aggregates \citep[see also][]{blum-et-al1998}. 
The rolling friction force between two spherical particles is given by \citep{dominik-tielens1995}
\begin{eqnarray}
F_\mathrm{roll}&=&6 \pi \gamma \xi_\mathrm{crit} . 
\end{eqnarray}
Therefore, we obtain
\begin{eqnarray}
F_\mathrm{roll}&=& 0.57~\mathrm{nN} \left({\frac{\gamma}{0.15~\mathrm{J~m}^{-2}}}\right)  \left({\frac{\xi_\mathrm{crit}}{0.2~\mathrm{nm}}}\right) , 
\end{eqnarray}
which coincides with the experimental results.\footnote{Although there is an uncertainty in the value of $\xi_\mathrm{crit}$, it should be on the order of interatomic distance \citep{dominik-tielens1995}.}
Microscopic analyses of the collision experiments similar to those of \citet{blum-wurm2000} were conducted by \citet{gundlach-et-al2011} to measure the rolling friction force on uncoated amorphous silica spheres.
They determined the rolling friction force of $F_\mathrm{roll} = 1.21 \pm 0.36~\mathrm{nN}$ for the gravitational restructuring of aggregates consisting of amorphous silica spheres of $a = 0.75~\micron$ in a dry nitrogen atmosphere.
Since we expect that the surface energy of $\gamma \approx 0.25~\mathrm{J~m}^{-2}$ is appropriate for uncoated amorphous silica, we may write 
\begin{eqnarray}
F_\mathrm{roll}&=& 0.94~\mathrm{nN} \left({\frac{\gamma}{0.25~\mathrm{J~m}^{-2}}}\right)  \left({\frac{\xi_\mathrm{crit}}{0.2~\mathrm{nm}}}\right) , 
\end{eqnarray}
which is in agreement with the measured value within the error bars.
Independently, \citet{heim-et-al1999} derived $F_\mathrm{roll} = 0.85 \pm 0.16~\mathrm{nN}$ from their AFM measurements of the force on chains of amorphous silica spheres with $a = 0.95~\micron$.
Although their AFM measurements were performed under ambient conditions, the contact areas of amorphous silica spheres seem to be established in a vacuum \citep[200~Pa; see][]{blum2000}.
Therefore, the measured rolling motion of amorphous silica spheres complies with the JKR theory, provided that the assumption of surface energy properly conforms to the circumstances.

\subsection{Coagulation Growth of Submicron Silicate Grains in Protoplanetary Disks}

In a protoplanetary disk, the formation of planetesimals proceeds with the coagulation of small grains, although one must resort to other formation mechanisms such as gravitational instabilities if they do not stick each other \citep{johansen-et-al2014}.
The coagulation of H$_2$O ice-coated grains of $a = 0.1~\micron$ could form planetesimals beyond the so-called snow line of the disk \citep{okuzumi-et-al2012}.
Inside the snow line, there has been a debate as to whether or not agglomerates of silicate grains could grow to planetesimals \citep[e.g.,][]{brauer-et-al2008}.
This is the most important issue for the formation of terrestrial planets, since the planets were most likely formed by an agglomeration of silicate grains rather than H$_2$O ice-coated grains. 
Silicate aggregates would grow to planetesimals by coagulation, only if the critical collision velocity $v_\mathrm{disrupt}$ exceeds a typical relative velocity between aggregates in protoplanetary disks.
Numerical calculations of relative velocities between aggregates in protoplanetary disks suggest that a typical collision velocity may reach approximately $50~\mathrm{m~s^{-1}}$ at 1\,AU from the central star \citep{weidenschilling-cuzzi1993,brauer-et-al2008}.
In contrast, previous numerical models have resulted in $v_\mathrm{disrupt} \la 6~\mathrm{m~s^{-1}}$ with the assumption of $\gamma = 0.025~\mathrm{J~m}^{-2}$ \citep{wada-et-al2009}.
Consequently, there is a common belief that it is hard for silicate aggregates to prevail over collisional destruction in protoplanetary disks.
In favor of $\gamma \approx 0.25~\mathrm{J~m}^{-2}$, however, we may write
\begin{eqnarray}
v_\mathrm{disrupt}
&\approx &54~\mathrm{m~s^{-1}} \left({\frac{\epsilon}{10}}\right) \left(\left\{\frac{1}{0.9711}\right\}-\left\{\frac{0.0289}{0.9711}\right\}\left\{\frac{\nu}{0.17}\right\}^{2}\right)^{1/3} \left(\frac{a}{0.1~\micron}\right)^{-5/6} \nonumber \\
&& \times {\left(\frac{\gamma}{0.25~\mathrm{J~m^{-2}}}\right)}^{5/6} {\left(\frac{E}{70~\mathrm{GPa}}\right)}^{-1/3} \left(\frac{\rho}{2.0\times{10}^{3}~\mathrm{kg~m^{-3}}}\right)^{-1/2} .
\end{eqnarray}
Therefore, we cannot rule out the possibility that agglomerates of amorphous silica particles with $a = 0.1~\micron$ grow to planetesimals by coagulation. 
Since the critical collision velocity as well as the surface energy for amorphous silica increases with temperature, the coagulation growth of silicate aggregates becomes easy in the vicinity of the star. 
In addition, quartz is expected to have a higher surface energy than amorphous silica, so that agglomerates of crystalline silicate grains may have higher critical collision velocities than those of amorphous silicate grains \citep[see][]{parks1984}.
In fact, \citet{poppe-et-al2000} have shown that the critical velocity for the sticking of enstatite grains is several times higher than that of amorphous silica grains onto the same target.
As a result, we assert that dust aggregates consisting of silicate grains with $a = 0.1~\micron$ could grow to planetesimals via coagulation in protoplanetary disks.

\subsection{Coagulation Growth of Submicron Silicate Grains in Molecular Clouds}

In the previous studies on dust coagulation in molecular clouds, a surface energy of $\gamma = 0.025~\mathrm{J~m}^{-2}$ has been a common assumption for silicate grains \citep{ormel-et-al2009,hirashita-li2013}.
This assumption led theorists to conclude that the collision velocity exceeds the critical velocity for sticking between bare silicate grains and thus the grains must be coated by H$_2$O ice mantles to coagulate in molecular clouds.
Nevertheless, that conclusion has now become obscured since the commonly assumed value of the surface energy is one order of magnitude lower than the most likely value for amorphous silica.
The critical velocity for sticking between identical elastic spheres of radius $a$ is given by \citep{chokshi-et-al1993,thornton-ning1998,brilliantov-et-al2007}
\begin{eqnarray}
v_\mathrm{stick}&=&{\left({\frac{27c_1\pi^{2/3}}{4}}\right)}^{1/2} {\left[{\frac{\gamma^5{\left({1-\nu^2}\right)}^2}{a^5\rho^3E^2}}\right]}^{1/6} . 
\label{critical-velocity}
\end{eqnarray}
We obtain the critical velocity for sticking between amorphous silica spheres of $a=0.1~\micron$ as
\begin{eqnarray}
v_\mathrm{stick}
&\approx &4.2~\mathrm{m~s^{-1}} {\left({\frac{c_1}{0.935}}\right)}^{1/2} \left(\left\{\frac{1}{0.9711}\right\}-\left\{\frac{0.0289}{0.9711}\right\}\left\{\frac{\nu}{0.17}\right\}^{2}\right)^{1/3} \left(\frac{a}{0.1~\micron}\right)^{-5/6} \nonumber \\
&& \times {\left(\frac{\gamma}{0.25~\mathrm{J~m^{-2}}}\right)}^{5/6} {\left(\frac{E}{70~\mathrm{GPa}}\right)}^{-1/3} \left(\frac{\rho}{2.0\times{10}^{3}~\mathrm{kg~m^{-3}}}\right)^{-1/2} .
\end{eqnarray}
A typical collision velocity between grains of $a = 0.1~\micron$ is estimated to be $8.3~\mathrm{m~s^{-1}}$ in turbulent molecular clouds at a molecular density of $n = 10^{11}~\mathrm{m}^{-3}$ \citep{ormel-et-al2009}.
Since the collision velocity of grains is proportional to $n^{-1/4}$, the critical velocity for sticking between amorphous silica spheres exceeds the collision velocity at $ n= 10^{13}~\mathrm{m}^{-3}$.
In summary, we confirm that amorphous silicate grains of $a = 0.1~\micron$ hardly grow by coagulation in a typical turbulent molecular cloud at $n = 10^{11}~\mathrm{m}^{-3}$, while they do grow easily in a dense core at $n \ga 10^{13}~\mathrm{m}^{-3}$.

\subsection{Effect of Grain Mantles on the Coagulation of Silicate Grains in Molecular Clouds}

Although amorphous silicate grains in the size range of submicrometers could stick together in dense cores, their surfaces are expected to accrete ice molecules, mainly amorphous H$_2$O ice.
It has been commonly accepted that H$_2$O ice mantles help submicron silicate grains to proceed with coagulation in molecular clouds \citep{ormel-et-al2009,hirashita-li2013}.
Note that the surface energy of $0.37~\mathrm{J~m^{-2}}$ is often assumed for H$_2$O ice in the majority of previous studies \citep{chokshi-et-al1993,dominik-tielens1997,ormel-et-al2009,hirashita-li2013}.
Recently, \citet{gundlach-blum2015} have derived the surface energy of $0.19~\mathrm{J~m^{-2}}$ for crystalline water ice from their collision experiments in vacuum.
Therefore, we do not consider that H$_2$O ice-coated silicate grains have higher efficiencies of coagulation growth than bare silicate grains in molecular clouds.
The reason for the refusal to accept the common belief is that the surface energy for amorphous H$_2$O ice is lower than that for amorphous silicate, although the Young's modulus of H$_2$O ice is one order of magnitude smaller than that of amorphous silica.
Because the surface energy for crystalline H$_2$O ice gives the upper limit to the surface energy for amorphous H$_2$O ice, which is most likely close to the surface tension for liquid water, we expect that the most realistic value is $\gamma \la 0.1~\mathrm{J~m^{-2}}$ for amorphous H$_2$O ice.
It is worth noting that ice mantles in a dense core are processed to form complex organic mantles, which have higher sticking efficiencies than silicates and ices \citep{kouchi-et-al2002,kimura-et-al2003}.
We can estimate the critical velocity for sticking between organic spheres of $a=0.1~\micron$ as follows:\footnote{For the sake of simplicity, we ignore the effect of the silicate core on the critical velocity for sticking, since it is a common practice to consider only the elastic properties of the mantle material \citep[e.g.,][]{ormel-et-al2009,hirashita-li2013}.}
\begin{eqnarray}
v_\mathrm{stick}
&\approx &66~\mathrm{m~s^{-1}} {\left({\frac{c_1}{0.935}}\right)}^{1/2} \left(\left\{\frac{1}{0.75}\right\}-\left\{\frac{0.25}{0.75}\right\}\left\{\frac{\nu}{0.5}\right\}^{2}\right)^{1/3} \left(\frac{a}{0.1~\micron}\right)^{-5/6} \nonumber \\
&& \times {\left(\frac{\gamma}{0.021~\mathrm{J~m^{-2}}}\right)}^{5/6} {\left(\frac{E}{0.09~\mathrm{MPa}}\right)}^{-1/3} \left(\frac{\rho}{0.96\times{10}^{3}~\mathrm{kg~m^{-3}}}\right)^{-1/2}, \label{vcr_or}
\end{eqnarray}
where the elastic properties are taken from polymers of high molecular mass and with a high fraction of dangling chains \citep{vaenkatesan-et-al2006}.
Accordingly, the most plausible route to coagulation of submicron amorphous silicate grains in molecular clouds is cohesion between organic materials covering the silicate grains.

\subsection{Aggregates of Silicate Nanoparticles in the Interstellar Medium}

It is worth emphasizing that the critical velocity for sticking decreases with the sizes of colliding particles, while the relative collision velocity increases.
The size dependence of Equation~(\ref{critical-velocity}) reveals that the critical velocities for sticking between amorphous silica spheres of $a \la 0.04~\micron$ exceed their collision velocities in a molecular cloud with a density of $n = 10^{11}~\mathrm{m}^{-3}$.
This indicates that amorphous silica nanoparticles could grow by coagulation to form aggregates in a typical turbulent molecular cloud without the coating of organic material.
Indeed, elemental analysis of crater-like features on the aluminum foils of the Stardust Interstellar Dust Collector suggests that submicrometer-sized silicate grains of interstellar origin are aggregates of silicate nanoparticles \citep{westphal-et-al2014}.
\citet{tielens-et-al1994} estimated that grain--grain collisions in interstellar shocks commence vaporization at a collision velocity above $19~\mathrm{km~s^{-1}}$ and shattering at $0.4~\mathrm{km~s^{-1}}$ for silicate grains.
Since our estimate of the critical velocity for disruption also results in $v_\mathrm{disrupt} \approx 0.4~\mathrm{km~s^{-1}}$ at $a = 0.01~\micron$, it is most likely that the Stardust interstellar dust samples of aggregates did not experience aggregate--aggregate collisions at velocities exceeding $0.4~\mathrm{km~s^{-1}}$.
Consequently, the Stardust mission has unveiled not only that the coagulation growth of silicate nanoparticles takes place in the interstellar medium, but also that the aggregates did not suffer from severe shattering by mutual collisions in the interstellar medium.

\acknowledgments

We would like to thank Masahiko Arakawa for useful discussions of the experimental conditions of measuring surface energies of amorphous silica particles.
H.K. is grateful to JSPS's Grants-in-Aid for Scientific Research (\#21340040, \#26400230).

\clearpage




\end{document}